# Fixed-Attention Mechanism for Deep-Learning-Assisted Design of High-Degree-of-Freedom 3D Metamaterials

HUANSHU ZHANG,[1,2] LEI KANG,[1] SAWYER D. CAMPBELL,[1] KAISHUN ZHANG,[2] DOUGLAS H. WERNER,[1,3] AND ZHAOLONG CAO[2,*]

[1]*Department of Electrical Engineering, The Pennsylvania State University, University Park, PA 16802, USA*
[2]*School of Electronics and Information Technology, Sun Yat-Sen University, Guangzhou 510275, China*
[3]*dhw@psu.edu*
[*]*caozhlong@mail.sysu.edu.cn*

**Abstract:** The traditional design approaches for high-degree-of-freedom metamaterials have been computationally intensive and, in many cases, even intractable due to the vast design space. In this work, we introduce a novel fixed-attention mechanism into a deep learning framework to address the computational challenges of metamaterial design. We consider a 3D plasmonic structure composed of gold nanorods characterized by geometric parameters and demonstrate that a Long Short-Term Memory network with a fixed-attention mechanism can improve the prediction accuracy by 48.09% compared to networks without attention. Additionally, we successfully apply this framework for the inverse design of plasmonic metamaterials. Our approach significantly reduces computational costs, opening the door for efficient real-time optimization of complex nanostructures.

## 1. Introduction

Metamaterials are artificial materials made of subwavelength structures that exhibit unique electromagnetic properties rarely found in nature [1–3], enabling functionalities such as negative refractive index [4], near-field focusing [5], and cloaking [6]. Designing these materials, however, is particularly challenging due to their performance typically depending on the resonant behavior of their small-scale structures, making it difficult to pinpoint the ideal design parameters for specific functions like achieving particular transmission or reflection spectra. Traditionally, metamaterial design processes have relied on physics-based numerical simulations (*e.g.*, full-wave Maxwell solvers) to predict and fine-tune their electromagnetic properties. Techniques such as the finite element method (FEM) [7,8] and finite difference time domain (FDTD) [9,10] algorithms are widely used, as they discretize Maxwell's equations to handle the complex geometries and materials ranging from antennas [11,12] and diffractive surfaces [13,14] to metamaterials [15,16] and photonic circuits [17,18]. Although essential, these simulations are computationally intensive, especially when navigating design spaces with high degrees of freedom (DoF). As design complexity increases, the scale of design spaces grows exponentially, requiring numerous simulation iterations that can overwhelm traditional methods [19,20]. Therefore, developing specialized algorithms to accelerate the design process is crucial for overcoming these limitations.

Machine learning, particularly deep learning, has emerged as a powerful tool to address the computational challenges of metamaterial design [21–23]. Compared to traditional approaches, Deep Neural Networks (DNNs) and other machine learning algorithms, once trained, have demonstrated multiple orders of magnitude evaluation speedups [24]. Various types of neural networks are suited for different tasks: fully connected layers are ideal for mapping nonlinear relationships between inputs and outputs, Convolutional Neural Networks (CNNs) excel in grid-like data processing (*e.g.*, images), and recurrent networks, such as Long Short-term Memory (LSTM) networks, are effective for sequence data [25]. Transformer [26] networks revolutionize tasks in natural language processing by leveraging self-attention mechanisms to capture global dependencies efficiently. DNNs have demonstrated the potential to approximate the electromagnetic behavior of metamaterials without directly solving Maxwell's equations [24], enabling rapid design iterations and real-time optimization.

Recent advancements in DNN-based models have proven their accuracy and efficiency in predicting the electromagnetic responses of a wide variety of metamaterials [27–30]. Two major approaches have emerged for metamaterial design using supervised machine learning [31]. The first approach involves pixelating a planar metamaterial, treating the pixelated representation as an image (often binary) and applying computer vision techniques such as CNNs [32–34] to optimize each pixel. This image-based method allows a high-DoF [20,35], as each pixel can be individually optimized to create intricate design configurations. However, this method is typically limited to the design of planar or 2.5D structures, where the height is an additional parameter. A true 3D metamaterial design using an image-based method remains a challenge, however, it offers significant benefits, such as the ability to realize intrinsic chiral structures. The second approach employs a parameter vector to describe a metamaterial structure, where a set of parameters defines the structure's geometry and material properties. Fully connected layers are often used in this context to predict spectral response of plasmonic systems [36–38], all-dielectric systems [39,40], and high-Q systems [41–43], including multi-layer thin films [44], cylinders [28], spheres [45], and nanofins [46]. Other neural network architectures have also been explored. For instance, Mao *et al.* have exploited Transformer networks to design trapezoidal 2D grating metasurfaces [47]. In addition, Deng *et al.* have used an LSTM neural network to predict the extinction ratio of a nanofin metasurface [48], and achieved a forward prediction mean-squared-error (MSE) of 0.08 on the valid set. However, these methods typically address designs with DoF fewer than eight, limiting their application to more complex structures due to the exponentially growing need for training data. Introducing more DoF can enable more flexibility in controlling electromagnetic waves with metamaterials [49]. Therefore, there is a clear need for faster and more efficient methods to design high-DoF 3D metamaterials, to unlock their full potential for practical applications.

Here, we present an LSTM network with a fixed-attention mechanism to address the challenges of designing fully 3D metamaterials with high-DoF. This mechanism improves the prediction accuracy of transmission spectra by 48.09% compared to networks without it. Unlike the widely known multihead self-attention mechanism used in transformers, which has been applied to metamaterial design in previous studies [47,50], we found that the fixed-attention mechanism performs better in high-DoF scenarios with limited training data. Specifically, the fixed-attention mechanism assigns static but optimized weights to input parameters, focusing on critical design variables during training. To demonstrate its efficacy, we applied this network to predict the performance of a plasmonic metamaterial composed of two gold nanorods embedded in a dielectric substrate, characterized by twelve parameters. The designed metamaterial can be fabricated layer-by-layer through repeating e-beam lithography patterning, metal evaporation, and lift-off processes with careful planarization. Detailed experimental protocols can be found in ref [51–53]. In addition to the high accuracy of the forward prediction, we further demonstrate the inverse design of this metamaterial using a tandem network approach, paving the way for the design of more complex and practical metamaterials.

## 2. Methods

*2.1 Plasmonic Meta-Atoms*

A schematic of the proposed plasmonic metamaterial structure is depicted in Fig. 1a and 1b. Each unit cell consists of two gold nanorods embedded in a dielectric substrate (n = 1.5), with each nanorod defined by six parameters: center position ($x$, $y$, $z$), length ($l$), width ($w$), and rotation angle ($\theta$) with respect to the z-axis. For two nanorods, their structure parameters can be represented by a parameter vector $v = (x_1, y_1, z_1, l_1, w_1, \theta_1, x_2, y_2, z_2, l_2, w_2, \theta_2)$. Each unit cell has a period $P_x=P_y=400\ nm$. The units of measurement are nanometers (for position and size) and degrees (for rotation). For practical considerations, when randomly generating the dataset, the *x*- and *y*-positions are limited between -170 and 170 nm with a step size of 10 nm, and *z*-positions are limited between -300 and 300 nm with the same step size. The length of the nanorods is selected from 60 to 300 nm, with a step size of 5 nm. The width of the nanorods is calculated based on a random aspect ratio $r = l/w$ of the nanorod, ranging from 2 to 10 with a step size of 0.2. The rotation angle is selected between -90° and 90°, with a 5° step size. The metamaterial is illuminated by a left-handed circularly polarized wave perpendicular to the structure.

The dataset was created via commercial full-wave electromagnetic simulation software (Lumerical FDTD Solutions) using randomly generated parameter vectors on a server with two Intel(R) Xeon(R) Gold 6258R CPUs and 1.5TB memory. The simulation domain dimensions were set to a size of 400 nm × 400 nm × 4000 nm in the *x*-, *y*-, and *z*-directions, respectively. Bloch boundary conditions were applied in the *x*- and *y*-directions, with a perfectly matched layer (PML) boundary condition in the *z*-direction. A mesh grid with a cell size of 3 nm × 3 nm × 2 nm along the *x*-, *y*-, and *z*-directions was employed. The transmission spectra were computed over a wavelength range of 700 to 1400 nm, discretized into 301 equally spaced points. A total of 6,493 data sets were generated, with 6,393 used for training and 100 reserved for testing. It is worth noting that our training set is extremely small, given that the design space of the two-rod system comprises $3.09 \times 10^{19}$ possible structures, making traditional iterative methods computationally prohibitive.

*2.2 Forward Networks*

We used a forward neural network to predict the transmission spectra based on the parameter vectors of the proposed structure, as shown in Fig. 1c. This network takes the parameter vector as input and predicts the corresponding transmission spectra. The architecture includes a fixed-attention mechanism layer, followed by four LSTM layers, each containing 1,024 units. A Tanh activation function is applied between each LSTM layer, while the sigmoid activation function is used after the output layer to constrain the predicted values between 0 and 1. All models were trained using the Adam optimizer with an initial learning rate of 0.001, which was reduced by 75% every 250 epochs. The neural network models were developed using Python 3.12.4 and the open-source PyTorch 2.3.1 framework with CUDA version 11.5. The network was trained on a machine with 4 NVIDIA GeForce RTX 2080 Ti GPUs, 1 Intel(R) Core(TM) i9-9960X CPU and 128GB memory. The overall performance is evaluated using the MSE metric:

$$MSE = \frac{1}{n}\sum_{k=1}^{n}(\widehat{T_k} - T_k)^2 \quad (1)$$

where $\widehat{T_k}$ is the *k*-th spectral value predicted by the network, and $T_k$ is the *k*-th ground truth value. To ensure consistent weight bias across each parameter, all input parameters are normalized according to the equation:

$$x_{i,normalized} = \frac{x_i - \bar{x}}{\sigma_x} \quad (2)$$

where $\bar{x}$ is the average value of the x-position coordinate across all 6,493 data points, and $\sigma_x$ is the standard deviation. For each element in the parameter vector, we performed this normalization with respect to its own average value and standard deviation.

The fixed-attention mechanism is implemented using a 12-by-12 matrix $M$ (Fig. 1d). This matrix is applied to the input vector through matrix multiplication, with each element in the attention matrix learnable and updated during backpropagation. After training, the matrix assigns optimized weights to the input parameters, effectively focusing the model's "attention" on specific elements. Before applying the attention matrix, each parameter is normalized to a common scale to ensure all input parameters share a comparable amplitude. The attention matrix then modifies these amplitudes, assigning more weight to certain parameters based on their importance to the prediction task. Notably, after training, the attention matrix assigns higher weights to the length and width parameters, as indicated by Fig. 1d. This aligns with the physical intuition that the transmission spectrum is primarily influenced by the aspect ratio of individual nanorods, while mutual interaction between rods, represented by the positions and rotation angles, plays a secondary role in the response to the external field and receives less weight. Unlike the self-attention mechanisms in Transformers, which dynamically adjust attention values based on input features, the fixed-attention mechanism assigns static but optimized weights during training. This approach simplifies the model while still enhancing its ability to capture critical design parameters, making it particularly effective in reducing the amount of training data required for complex metamaterial designs.

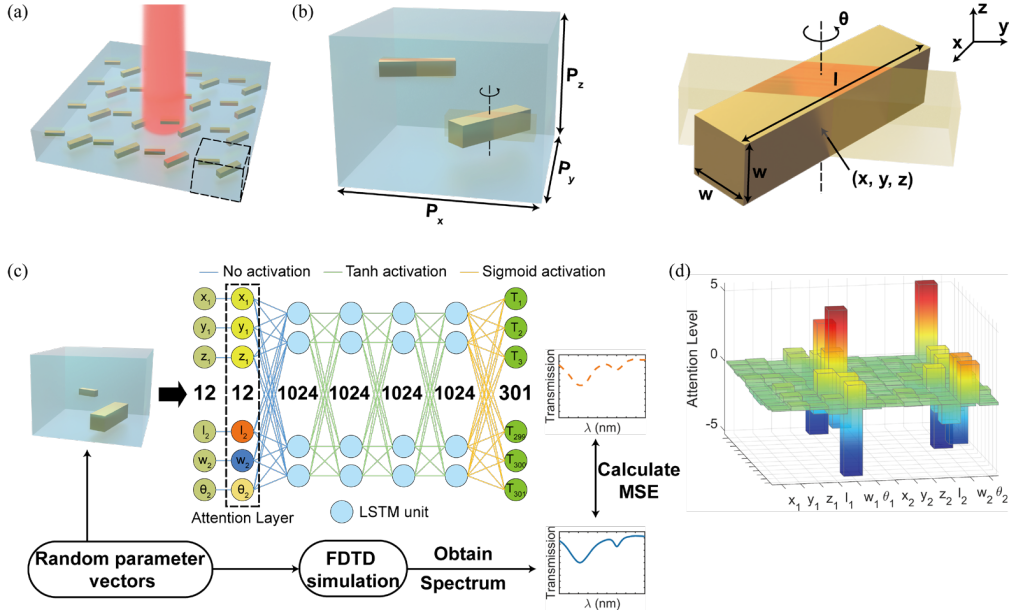

**Fig. 1.** (a) The structure of the plasmonic metamaterial. (b) A unit cell that contains two gold nanorods embedded in a dielectric substrate, with each gold nanorod defined by six parameters (positions, sizes, orientation) resulting in a total of twelve parameters to describe the configuration. (c) Schematic of the forward network architecture, including an attention layer, four LSTM layers with 1,024 units each, and an output layer. The network takes the parameter vector as input and outputs the corresponding transmission spectrum ($T$). (d) Visualization of the fixed-attention matrix applied to the input vector, where the output is computed as the product of the parameter vector and the transposed attention matrix ($v' = v \times M^T$). Higher weights are assigned to the length ($l_1$, $l_2$) and width ($w_1$, $w_2$) parameters, as shown by the taller bars, reflecting their greater influence on the transmission spectrum.

## 3. Results and discussion

We compared our fixed-attention LSTM network with other network configurations and mechanisms to evaluate its effectiveness. Fig. 2a plots the learning curves of the model with a fixed-attention mechanism against the model without it. Neither case shows signs of overfitting, indicating that the regression problem was well-handled. We also assessed performance across different network configurations by evaluating their validation MSE, as shown in Fig. 2b. The configurations include various attention mechanisms (fixed attention, no attention, and multihead attention [26]), network architectures (LSTM and DNN), and activation functions (ReLU and Tanh). These results indicate that the LSTM network with a fixed-attention mechanism and Tanh activation function outperforms the other configurations, achieving an MSE of $2.17 \times 10^{-3}$, nearly twice as accurate as the same network without attention (MSE = $4.18 \times 10^{-3}$). In contrast, applying fixed-attention to DNNs yielded only marginal performance gains, suggesting that this mechanism is less effective for fully-connected network performance than for LSTM networks. Interestingly, the multihead attention mechanism (3 attention heads with embedding dimension of 12) performed significantly worse across all network configurations, likely due to the limited dataset size. Multihead attention layers introduce additional parameters and increase model complexity. With a limited dataset, these extra parameters may not be adequately constrained, leading to overfitting. In contrast, simpler network architectures often incorporate stronger inductive biases, enabling them to generalize more effectively in data-scarce situations. This suggests that multihead attention, which excels at capturing global dependencies, may be less effective for metamaterial design tasks where local dependencies (such as aspect ratio and orientation of individual nanorods) dominate the electromagnetic response. The superior performance of the fixed-attention LSTM model suggests that the mechanism efficiently captures non-linear relationships between the design parameters and the transmission spectrum, making it an ideal approach for high-dimensional metamaterial designs, where accurate predictions are critical for real-time applications in nanophotonics. In addition, we evaluated our model using alternative dataset splits, ranging from a split with 100 test samples and 6,394 training samples to one with 649 test samples, 649 validation samples, and 5,195 training samples (a typical 4:1 split), and found that the MSE values remained consistently comparable, varying from $2.17 \times 10^{-3}$ to $3.48 \times 10^{-3}$.

Fig. 2c illustrates the predicted transmission responses (with and without attention) compared to direct FDTD simulation results. The predictions from the network with the fixed-attention mechanism (orange dashed curve) closely align with the simulated results (blue curve), effectively resolving resonant peaks, which are critical features in the transmission spectra. By contrast, the network without the fixed-attention mechanism (red dashed curve) deviates noticeably from the FDTD simulation results, particularly in regions with sharp spectral features. This performance improvement is quantitatively reflected in the MSE values. After training for 450 epochs, the network without attention achieved MSE values of $4.15 \times 10^{-3}$ and $4.18 \times 10^{-3}$ for the training and test set, respectively, while the network with fixed-attention achieved significantly lower MSE values of $9.70 \times 10^{-4}$ and $2.17 \times 10^{-3}$ for the training and test set, representing a 48.09% performance improvement on the test set. Furthermore, calculating the entire set of structural parameters required only approximately 3 ms through the deep neural network, compared to about 50 minutes per case using FDTD simulation.

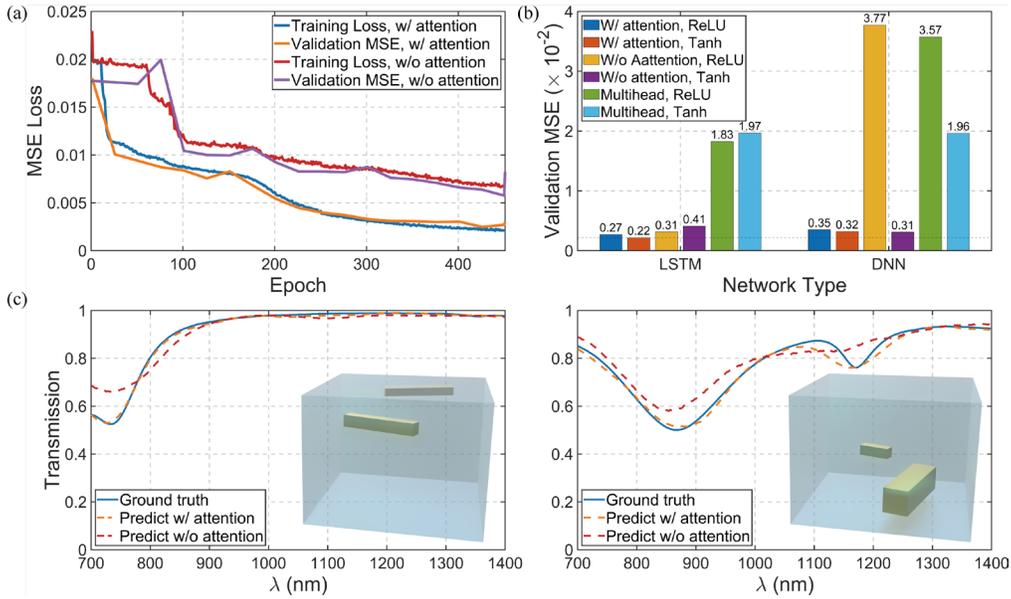

**Fig. 2.** (a) Learning curves for models with and without the fixed-attention mechanism. (b) Comparison of validation MSE across different attention mechanisms (fixed-attention, no attention, multihead attention), different network types (LSTM and DNN) and different activation function types (ReLU, Tanh). (c) Predicted and simulated transmission spectra for two parameter vectors from the test set. The left figure shows a sample with one resonance peak, and the right figure shows a sample with two resonance peaks. The parameter vectors for these samples in the left figure are [-70, -30, 270, 170, 19, 20, 100, 80, 120, 190, 30, -20], and in the right figure are [-120, 50, -200, 220, 61, 75, 60, 50, -20, 75, 25, -20]. The corresponding MSE values in the left figure are $3.71 \times 10^{-5}$ (with attention) and $1.68 \times 10^{-3}$ (without attention), and for the right figure are $4.04 \times 10^{-4}$ (with attention) and $2.22 \times 10^{-3}$ (without attention).

To evaluate our model's ability for handling complex metamaterials with a high-DoF, we compared our design approach with those in reported studies on AI-assisted 3D metamaterial designs characterized by parameter vectors in Table 1. We defined normalized mean squared error (NMSE) as the ratio of the mean squared error to the square of the prediction range. Using a comparable training set size, our approach operates within a significantly larger design space while achieving NMSE values similar to those reported. This demonstrates our model's superior efficiency in managing high-dimensional metamaterial designs.

**Table 1. Comparison Across Recent Seq2Seq Network-Assisted Metamaterial Design Studies.**

| Work | Number of Free Parameters | Design Space | Training Set Size | NMSE |
|---|---|---|---|---|
| [40] | 3 | $1.41 \times 10^{8}$ * | 3,157 | $9.4 \times 10^{-4}$ |
| [54] | 5 | $4.74 \times 10^{6}$ * | 7,000 | $1.62 \times 10^{-10}$ |
| [55] | 5 | $3.51 \times 10^{9}$ * | 25,000 | $2 \times 10^{-4}$ |
| [39] | 8 | $8.157 \times 10^{8}$ | 18,000 | $1.6 \times 10^{-3}$ |
| [37] | 5 | $4.25 \times 10^{7}$ * | 23,000 | $4 \times 10^{-5}$ |

| Ref | | | | |
|---|---|---|---|---|
| [56] | 4 | 4.96×10$^8$ * | 30,000 | 8.7×10$^{-3}$ |
| [57] | 4 | 9.28×10$^8$ * | 36,000 | 6.7×10$^{-4}$ |
| [58] | 5 | 1.05×10$^{10}$ * | 8,400 | 1.3×10$^{-4}$ |
| [59] | 4 | 2.83×10$^6$ | 320 | 1.2×10$^{-2}$ |
| [60] | 5 | 2.60×10$^9$ * | 18,144 | 2.8×10$^{-4}$ |
| [61] | 5 | 1.65×10$^9$ * | 6,318 | 1.20×10$^{-4}$ |
| [62] | 6 | 1.97×10$^{14}$ | 4,812 | 2.66×10$^{-3}$ |
| [63] | 7 | 2.18×10$^7$ | 8,000 | 1.60×10$^{-4}$ |
| This work | 12 | 3.09×10$^{19}$ | 6,393 | 2.17×10$^{-3}$ |

**(\* indicates estimates based on parameter ranges provided in the respective studies.)**

### 2.3 Inverse Network

As a final example, we demonstrate the ability of our network in the inverse design of a metamaterial with a user-defined spectrum. Inverse design presents a more complex challenge compared to the forward problem, as multiple parameter vectors can yield similar optical responses, resulting in convergent issues during neural network training. To address this issue, we implemented a tandem network [44], which combines a forward network with a backward network, as illustrated in Fig. 3a. The training process is shown on Fig. 3b. After 20,000 epochs, the MSE reached $2.37 \times 10^{-3}$ and $1.86 \times 10^{-3}$ for training and test sets, respectively, indicating that training converged with no signs of overfitting. To illustrate the inverse network's performance, two inverse-designed results from the test set are plotted in Fig. 3c. The MSE values for these two results are $1.27 \times 10^{-2}$ (upper) and $1.18 \times 10^{-2}$ (lower). Although mismatches are observed between the target spectra and the predicted spectra, the MSE values are acceptable when compared with previous works. This mismatch between designed spectrum and target spectrum may come from the coupling effect between gold nanorods, as the pre-trained forward network assigns higher attention to the length and width of nanorods. This neglect of interaction may influence the accuracy of the inverse network. Hence, these results confirm the utility of both the forward and inverse networks as effective pre-calculation tools.

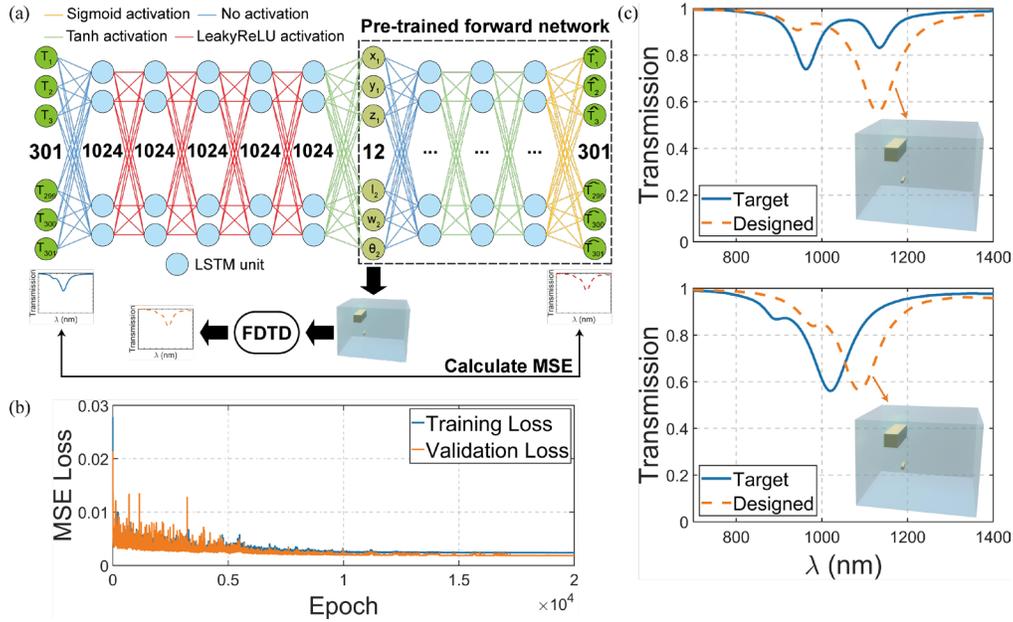

**Fig. 3.** (a) Structure of the tandem neural network. The training aims to accurately reproduce the desired input spectra. During the training of the backward network, the weight values in the pre-trained forward network remain unchanged. The pre-trained forward network has the same structure as shown in Fig. 1(c). (b) Learning curve for the backward training process. (c) Comparison between the designed optical responses (FDTD used to simulate the parameter vector predicted by backward network) and target responses from the test set. Designed parameter vectors in the upper figure are [169, 164, 320, 179, 46, -91, 165, -165, -327, 60, 10, -90], and in the lower figure are [169, 163, 319, 183, 46, -91, 165, -165, -327, 68, 13, -85].

## 4. Conclusion

In this work, we leveraged a neural network model with a fixed-attention mechanism that not only predicted the optical responses of high-DoF metamaterials but also provided spectral generalizability. The fixed-attention mechanism was applied in both forward and inverse design tasks, effectively addressing the non-convergence problem. This fixed-attention approach can be readily applied to any electromagnetic structures that are described by a parameter vector, providing a generalizable solution for the efficient design of complex metamaterials. We envision that the proposed method can be extended to the efficient design of multifunctional metadevices based on sophisticated plasmonic and/or dielectric structures.

**Data availability.** Data underlying the results presented in this paper are not publicly available at this time but may be obtained from the authors upon reasonable request.

**Acknowledgment.** The authors acknowledge Yumeng Qin for valuable discussions.

**Disclosures.** The authors declare no conflicts of interest.

**Funding.** National Key Basic Research Program of China (grant no. 2022YFA1206600); the National Natural Science Foundation of China (grant no. 61905290).